\pdfoutput=1
\documentclass[10pt]{iopart}
\usepackage{xcolor}
\usepackage{graphicx}
\usepackage{latexsym}
\usepackage{url}
\usepackage{xcolor}

\begin{document}
\def\aap{A\&A\,  }
\def \aapr{The Astronomy and Astrophysics Review}
\def\aaps{A\&AS  }
\def\acp{Anal. Cell. Pathol. } 
\def\aj{AJ  }
\def\aplett{Astrophys. Lett.\,  }
\def\apj{ApJ\,  }
\def\apjl{ApJ\,  }
\def\apjs{ApJS  }
\def\apss{Astrophysics and Space Science  }
\def\araa{ARA\&A  }
\def\azh{AZh}
\def\bain{BAN  }
\def\baas{Bulletin of the American Astronomical Society}
\def\cjaa{Chinese Astronomy and Astrophysics}
\def\eup{Europhys. Lett.  }
\def\fcp{Fundamentals of Cosmic Physics}
\def\iaucirc{IAU circ.  } 
\def\icarus{Icarus} 
\def\jaa{J. Astrophys. Astr.  }
\def\jpc{J. Phys. C  } 
\def\JPG{J. Phys. G\,  }
\def\jsp{J. Stat. Phys  } 
\def\jcap{Journal of Cosmology and Astroparticle Physic  } 
\def\jcp{J. Comput. Phys.  } 
\def\jcpp{Journal of Chemical  Physics  } 
\def\jrasc{JRASC  } 
\def\mnras{MNRAS\,  }

\def\na {New Astronomy\,  }
\def\nat{Nature\,  }
\def\npb{Nuc. Phys. B   }
\def\oe{Optic  Express }
\def\pasj{PASJ\,  }
\def\solphys{Sol. Phys.\,  }
\def\planss{Planet. Space Sci.  }
\def\pasp{PASP  }
\def\pasa{PASA  }
\def\POF{Physics of Fluids  }
\def\physrep{Phys. Rep.\,  }
\def\pla{Phys. Lett. A   }
\def\pra{Phys. Rev. A   }
\def\prb{Phys. Rev. B   }
\def\prc{Phys. Rev. C   }
\def\prd{Phys. Rev. D   }
\def\pre{Phys. Rev. E   }
\def\prl{Phys. Rev. Lett.    }
\def\physa{Phys. A    }
\def\rmp{Rev. Mod. Phys.  }
\def\rmxaa{Revista Mexicana de Astronomia y Astrofisica} 
\def\skytel{Sky and Telescope}
\def\ssr{Space Science Reviews} 
\def\rpp{Rep.Prog.Phys.   }
\def\sovast{Soviet Astronomy} 
\def\za{Z. Astrophys.  } 
\def\zap{Zeitschrift fur Astrophysik} 
\def\h0units{\mathrm{km\,s^{-1}\,Mpc^{-1}}}
\def\cunits{\mathrm{km\,s^{-1}}}
\def\lunits{\mathrm{erg\,s^{-1}}}
\newcommand{\om}{\Omega_{\rm M}}
\newcommand{\ok}{\Omega_K}
\newcommand{\ola}{\Omega_{\Lambda}}
\newcommand{\dl}{d_{\rm{L}}}
\newcommand{\mstar}{\ensuremath{m_{B}^\star}\,}
\def\sun{\hbox{$\odot$}}

\title
{
A new analytical solution for the distance modulus  in flat cosmology
}
\vspace{2pc}
\author     {Lorenzo  Zaninetti}
\address    {Physics Department,
 via P.Giuria 1,\\ I-10125 Turin,Italy }
\ead {zaninetti@ph.unito.it}

\begin {abstract}
A new analytical solution for the luminosity distance
in flat $\Lambda$CDM cosmology is derived  in terms 
of elliptical integrals of first kind with real 
argument.
The consequent derivation of the distance modulus 
allows evaluating   the Hubble constant, 
$H_0=69.77\pm 0.33$, 
$\om= 0.295\pm0.008$
and
the cosmological constant,  $\Lambda=
(1.194 \pm 0.017)10^{-52} \frac{1}{m^2}$.
\end{abstract}

\vspace{2pc}

\maketitle

\section{Introduction}

The release of two  catalogs  for  the distance modulus
of Supernova (SN)   of type Ia, namely, the Union 2.1 compilation, 
see \cite{Suzuki2012},
and    
the joint light-curve analysis (JLA), see \cite{Betoule2014},
allows matching the 
observed distance modulus  with 
the theoretical distance modulus of various cosmologies.
In this fitting procedure, the cosmological parameters are 
derived in a scientific and reproducible way.

We now focus our attention   on the flat 
Friedmann-Lema\^{i}tre-Robertson-Walker (flat-FLRW) cosmology.
A first fitting formula  has been
derived  by \cite{Pen1999}
and 
an approximate solution in terms 
of Pad\'e  approximant
has been introduced by \cite{Adachi2012}.
The presence 
of the elliptical integrals of the first kind
in the integral for  the luminosity distance in flat-FLRW cosmology
has been noted by 
\cite{Eisenstein1997,Liu2011,Meszaros2013}.
As a practical  example  
the luminosity
distance can be expanded into a series of orthonormal
functions and   
the two  cosmological parameters
turn out to be  
$H_0=70.43 \pm 0.33$  and $\om= 0.297\pm0.002$,
see \cite{Benitez-Herrera2013}.
This  paper first introduces in Section \ref{preliminaries} 
a 
framework useful to build  a new solution
for the luminosity distance in flat-FLRW cosmology, which 
will be derived in Section \ref{newflat}.

\section{Preliminaries}

\label{preliminaries}
This section  reviews 
the adopted statistical framework,
the $\Lambda$CDM cosmology,
and an existing solution for the luminosity distance
in flat-FLRW cosmology.

\subsection{The adopted statistics}

In the case of the distance modulus, 
the  merit function $\chi^2$ is
\begin{equation}
\chi^2  = \sum_{i=1}^N \biggr [\frac{(m-M)_i - (m-M)(z_i)_{th}}{\sigma_i}\biggl] ^2
\quad ,
\label{chisquare}
\end{equation}
where $N$ is the number of SNs, 
$(m-M)_i$ is the observed distance modulus
evaluated at redshift $z_i$,
$\sigma_i$ is the error in the observed distance
modulus evaluated at $z_i$,
and
$(m-M)(z_i)_{th}$ is the theoretical distance modulus 
evaluated at $z_i$, see formula (15.5.5) in \cite{press}.
The reduced  merit function $\chi_{red}^2$
is  
\begin{equation}
\chi_{red}^2 = \chi^2/NF
\quad,
\label{chisquarereduced}
\end{equation}
where $NF=N-k$ is the number of degrees  of freedom,
       $N$     is the number of SNs,
and    $k$     is the number of parameters.
Another useful statistical parameter is  the associated $Q$-value,
which has   to be understood as the
 maximum probability of obtaining a better fitting,
 see formula (15.2.12) in \cite {press}:
\begin{equation}
Q=1- GAMMQ (\frac{N-k}{2},\frac{\chi^2}{2} )
\quad ,
\end{equation}
where GAMMQ is a subroutine  for the incomplete gamma function.

The    goodness of the approximation 
in evaluating a physical 
variable  $p$  is evaluated by the percentage 
error
$\delta$ 
\begin{equation}
\delta = \frac{\big | p - p_{approx} \big |}
{p} \times 100
\quad ,
\label{error100}
\end{equation}
where $p_{approx}$ is an approximation of $p$.

\subsection{The standard cosmology}

We follow  \cite{Hogg1999},
where
the {\em Hubble
distance\/} $D_{\rm H}$
is defined as
\begin{equation}
\label{eq:dh}
D_{\rm H}\equiv\frac{c}{H_0}
\quad .
\end{equation}
The first parameter is 
 $\om$
\begin{equation}
\om = \frac{8\pi\,G\,\rho_0}{3\,H_0^2}
\quad ,
\end{equation}
where $G$ is the Newtonian gravitational constant and
$\rho_0$ is the mass density at the present time.
The second parameter is $\ola$
\begin{equation}
\ola\equiv\frac{\Lambda\,c^2}{3\,H_0^2}
\quad ,
\end{equation}
where $\Lambda$ is the cosmological constant,
see \cite{Peebles1993}.
These two parameters are connected with the
curvature $\ok$ by
\begin{equation}
\om+\ola+\ok= 1
\quad .
\end{equation}
The  comoving distance, $D_{\rm C}$,  is
\begin{equation}
D_{\rm C} = D_{\rm H}\,\int_0^z\frac{dz'}{E(z')}
\label{integralez}
\end{equation}
where $E(z)$ is the `Hubble function'
\begin{equation}
\label{eq:ez}
E(z) = \sqrt{\om\,(1+z)^3+\ok\,(1+z)^2+\ola}
\quad .
\end{equation}
The above integral does not have an analytical
solution  but  a solution 
in terms of Pad\'e  approximant  has been found,
see \cite{Zaninetti2016a}.

\subsection{A first formula for a flat-FLRW universe} 

The first model  starts from equation  (2.1) in \cite{Adachi2012} for
the luminosity distance, $\dl$, 
\begin{equation}
  \dl(z;c,H_0,\om) = \frac{c}{H_0} (1+z) \int_{\frac{1}{1+z}}^1
  \frac{da}{\sqrt{\om a + (1-\om) a^4}} \quad ,
  \label{lumdistflat}
\end{equation}
where $H_0$
is the Hubble constant expressed in     $\h0units$,
$c$ is the speed  of light expressed in $\cunits$,
$z$ is the redshift and 
$a$ is the scale-factor.
The indefinite integral, $\Phi(a)$, is 
\begin{equation}
\Phi(a,\om) = \int
  \frac{da}{\sqrt{\om a + (1-\om) a^4}}
\quad .
\end{equation}
The solution is in terms of
$\mathop{F\/}$, the Legendre integral or incomplete
elliptic integral of the first kind,
and is given in \cite{Zaninetti2016b}.

The luminosity distance is
\begin{equation}
  \dl(z; c,H_0,\om) =
\Re \bigg ( \frac{c}{H_0} (1+z) \big( \Phi(1) -\Phi(\frac{1}{1+z})
\big) \bigg) \quad ,
\end{equation}
where $\Re$ means the  real part.
The distance modulus is
\begin{equation}
(m-M) =25 +5 \log_{10}\bigg ( \dl(z; c,H_0,\om) \bigg)
\quad .
\label{distmod_elliptic}
\end{equation}

\section{A new formula for a flat-FLRW universe}

\label{newflat}
The second model for the flat cosmology starts 
from equation (1) for the luminosity 
distance in \cite{Baes2017} 
\begin{equation}
  \dl(z;c,H_0,\om) = \frac{c(1+z)}{H_0} \int_{0}^{z}\!{\frac {1}{\sqrt {{\it
\om}\, \left( 1+t \right) ^{3}
+1-{\it \om}}}}\,{\rm d}t
\quad .
\end{equation}
The above formula can be obtained from formula 
(\ref{integralez}) for the comoving distance 
inserting  $\ok=0$ and the variable of integration, $t$, 
denotes the redshift.

A first 
change in the parameter $\om$ introduces
\begin{equation}
s=\sqrt [3]{{\frac {1-{\it \om}}{{\it \om}}}}
\quad 
\label{sdef}  
\end{equation}
and the luminosity distance becomes
\begin{equation}
  \dl(z;c,H_0,s) =\frac{1}{H_0}  
c \left( 1+z \right) \int_{0}^{z}\!{\frac {1}{\sqrt {{\frac { \left( 1
+t \right) ^{3}}{{s}^{3}+1}}+1- \left( {s}^{3}+1 \right) ^{-1}}}}
\,{\rm d}t
\quad .
\end{equation}
The following change of variable, $t=\frac{s-u}{u}$, is performed 
for the luminosity distance which becomes 
\begin{eqnarray}
  \dl(z;c,H_0,s) =
\nonumber  \\
-\frac{c}{{\it H_0}\,{s}^{2}}  
 \left( 1+z \right)  \left( {s}^{3}+1 \right) \int_{s}^{{\frac {s}{1
+z}}}\!{\frac {u}{{u}^{3}+1}\sqrt {{\frac {{s}^{3} \left( {u}^{3}+1
 \right) }{{u}^{3} \left( {s}^{3}+1 \right) }}}}\,{\rm d}u
\quad .
\end{eqnarray}

Up to now we have  followed 
\cite{Baes2017} 
which continues introducing a new function $T(x)$;
conversely  we work directly on the resulting integral
for the luminosity distance:
which is  
\begin{eqnarray}
  \dl(z;c,H_0,s) =
-1/3\,{\frac {c  ( 1+z   ) {3}^{3/4}\sqrt {{s}^{3}+1}}{\sqrt {
s}{\it H_0}}
} \times \nonumber \\
{
\Bigg  ( {\it F}  ( 2\,{\frac {\sqrt {s  ( s
+1+z   ) }\sqrt [4]{3}}{s\sqrt {3}+s+z+1}},1/4\,\sqrt {2}\sqrt {3}
+1/4\,\sqrt {2}   ) 
}
\nonumber \\
{
-{\it F}  ( 2\,{\frac {\sqrt [4]{3
}\sqrt {s  ( s+1   ) }}{s+1+s\sqrt {3}}},1/4\,\sqrt {2}\sqrt {
3}+1/4\,\sqrt {2}   )   \Bigg ) }
\quad ,
\label{distlumflatnew}
\end{eqnarray}
where $s$ is given by Eq. (\ref{sdef}) and 
$F\left(\phi,k\right)$
is Legendre’s incomplete elliptic integral of the first kind,
\begin{equation}
F\left(\phi,k\right)=
\int_{0}^{\sin\phi}\frac{\mathrm{d}t}{\sqrt{1-t^{2}}\sqrt{1-k^{2}t^{2}}}
\quad ,
\end{equation}
see  \cite{NIST2010}.
The distance modulus is
\begin{equation}
(m-M) =25 +5 \log_{10}\bigg ( \dl(z; c,H_0,s) \bigg)
\quad ,
\label{distmod_elliptic_new}
\end{equation}
and therefore 
\begin{eqnarray}
(m-M)= 25
\nonumber \\
+5\,{\frac {1}{\ln  \left( 10 \right) }\ln  \left( -\frac{1}{3}\,{\frac {c
 \left( 1+z \right) {3}^{3/4} \left( {\it F_1}-{\it F_2} \right) 
\sqrt {{s}^{3}+1}}{\sqrt {s}{\it H_0}}} \right) }
\quad ,
\label{distmodflatnew}
\end{eqnarray}
where 
\begin{equation}
F_1=
{\it F} \left( 2\,{\frac {\sqrt {s \left( s+1+z \right) }
\sqrt [4]{3}}{s\sqrt {3}+s+z+1}},1/4\,\sqrt {2}\sqrt {3}+1/4\,\sqrt {2
} \right)
\end{equation}
and
\begin{equation}
F_2=
{\it F} \left( 2\,{\frac {\sqrt [4]{3}\sqrt {s \left( s+1
 \right) }}{s+1+s\sqrt {3}}},1/4\,\sqrt {2}\sqrt {3}+1/4\,\sqrt {2}
 \right) 
\quad , 
\end{equation}
with $s$ as defined  by Eq. (\ref{sdef}).

\subsection{Data Analysis}

\begin{figure}
\includegraphics[width=10cm,angle=-90]{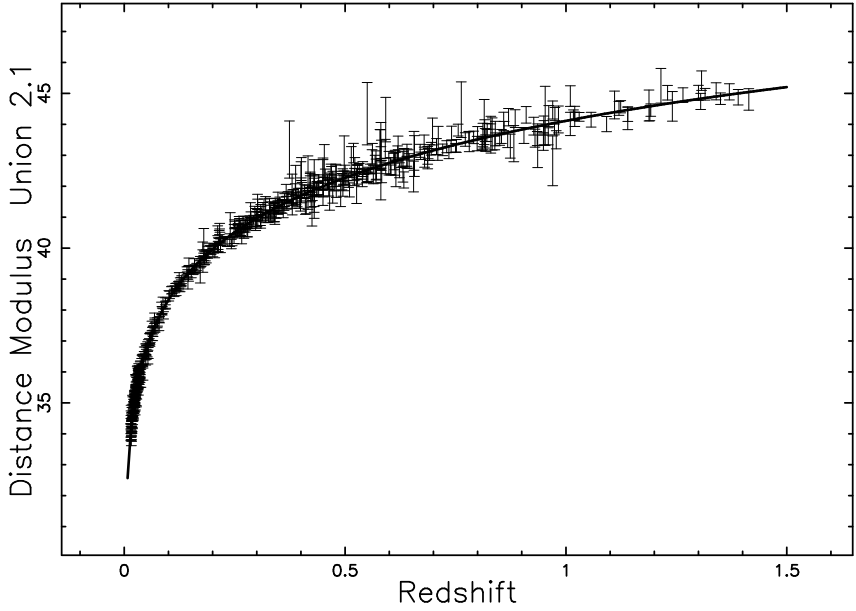}
\caption{
Hubble diagram for the  Union 2.1  compilation.
The solid line represents the best fit
for the exact distance modulus  in flat-FLRW cosmology 
as represented by Eq.~(\ref{distmodflatnew}).
Parameters as in first line of Table 
\ref{chi2valueflatmnras}; Union 2.1  compilation.
}
\label{distmunion21}
\end{figure}

In recent years, the extraction of the cosmological 
parameters from the distance modulus of SNs has become
a common practice, see among others 
\cite{Benitez-Herrera2013,Gupta2018,Jones2018}.
The best fit to the distance modulus
of SNs is here obtained by implementing  the
Levenberg--Marquardt  method
(subroutine MRQMIN in \cite{press}).
This method requires the fitting function, in our 
case Eq.~(\ref{distmodflatnew}), as well 
the first derivative $\frac{\partial (m-M)}{\partial H_0} $,
which has a simple expression, and
the first derivative $\frac{\partial (m-M)}{\partial \om} $,
which has a complicated expression.
A simplification can be introduced  by imposing
a fiducial value for the Hubble constant,
namely $H_0=70 \h0units$, see \cite{Ben-Dayan2013,Betoule2014}. 
We call this  model 
`flat-FLRW-1', where the `1' stands for there being one  parameter.
Table \ref{chi2valueflatmnras} presents $H_0$ and $\om$
for  the  Union 2.1  compilation of SNs  and
Figure \ref{distmunion21} 
displays the best  fit.
The reading of 
 this table  allows
to evaluate the goodness of the approximation, 
see (\ref{error100}), in the derivation
of  the Hubble constant in going from the supposed
true value  ($H_0=70 \h0units$)
to the deduced value ($H_0=69.77\,\h0units$),
which is $\delta=99.67\,\%$.
The JLA  compilation is available at
the Strasbourg Astronomical Data Centre (CDS),
and consists of 740 type I-a  SNs  for which
we have  the heliocentric redshift, $z$, the apparent
magnitude $\mstar$ in the B band, the error in $\mstar$, $\sigma_{\mstar}$,
the parameter $X1$,   the error in $X1$,
$\sigma_{X1}$,
the parameter $C$, the error in $C$, $\sigma_C$,  and
$\log_{10} (M_{stellar})$.
The observed distance modulus  is defined by
equation~(4) in \cite{Betoule2014}:
\begin{equation}
m-M =
-C\beta+{\it X1}\,\alpha-M_{{b}}+ \mstar
\quad.
\end{equation}
The  adopted parameters are
$\alpha=0.141$, $\beta=3.101$   and
\begin{equation}
M_{{b}} =  \left\{
\begin{array}{ll}
-19.05  & if~  M_{stellar} <    10^{10} M_{\sun} \\
-19.12  & if~  M_{stellar} \geq 10^{10} M_{\sun}
 \end{array}
\right.
\end{equation}
see line 1  in Table 10 of
\cite{Betoule2014}.
The uncertainty  in the observed distance modulus,
$\sigma_{m-M}$,
is  found by implementing the error
propagation equation (often called the law of errors of Gauss) when
the covariant terms are neglected, see  equation (3.14)
in \cite{bevington2003},
\begin{equation}
\sigma_{m-M} =
\sqrt {{\alpha}^{2}{\sigma_{{{\it X1}}}}^{2}+{\beta}^{2}{\sigma_{{C}}}
^{2}+{\sigma_{{{\it \mstar}}}}^{2}}
\quad .
\end{equation}

\begin{figure}
\includegraphics[width=10cm,angle=-90]{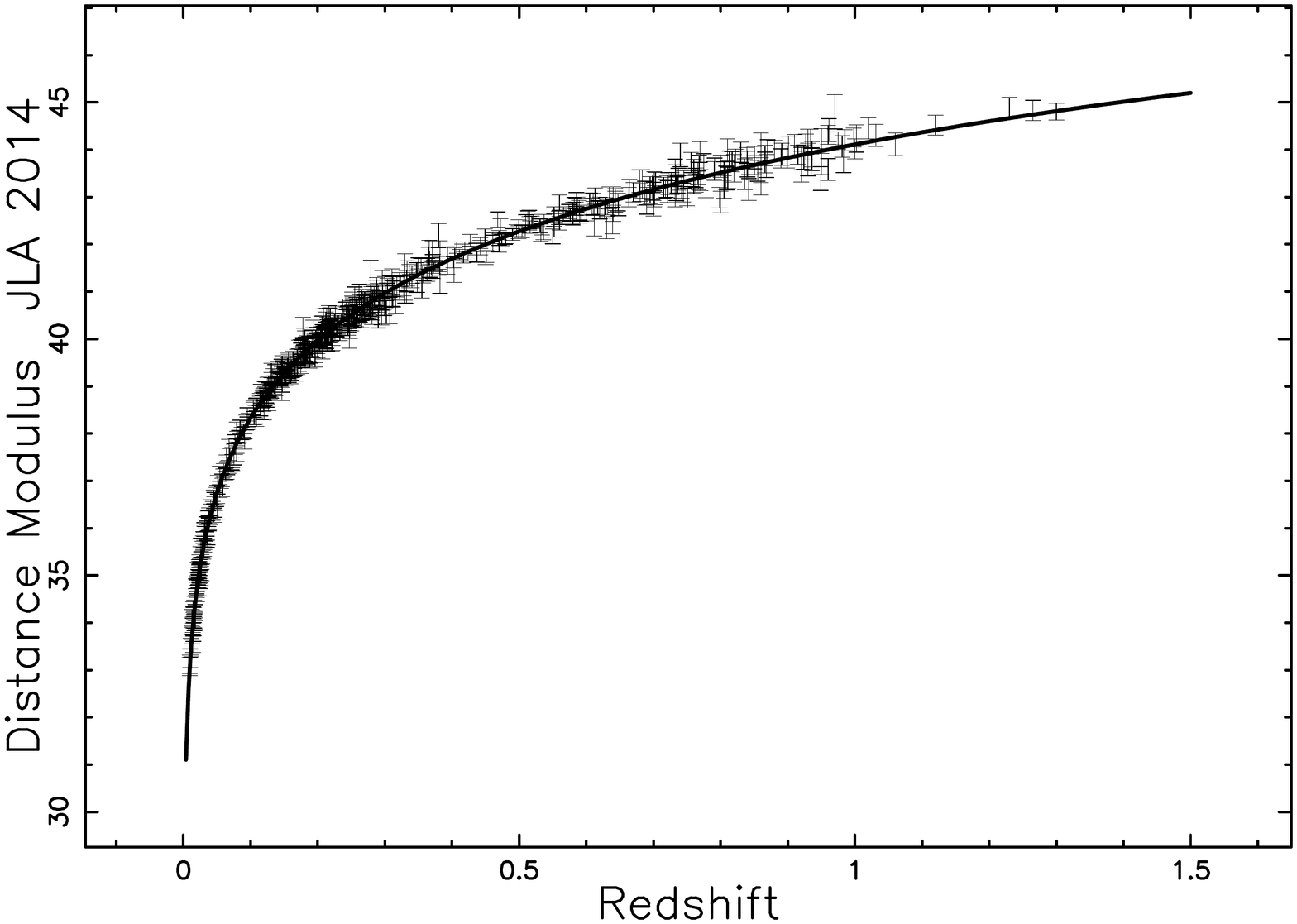}
\caption{
Hubble diagram for the  JLA  compilation.
The solid line represents the best fit
for the exact distance modulus  in flat-FLRW cosmology 
as represented by Eq.~(\ref{distmodflatnew}).
Parameters as in first line of Table 
\ref{chi2valueflatjla}.
}
\label{distmjla}
\end{figure}

The  parameters  as derived from the JLA compilation are 
presented  in Table \ref{chi2valueflatjla} 
and the fit is presented in Figure \ref{distmjla}.
\begin{table}[ht!]
\caption
{
Numerical values 
from  the  Union 2.1  compilation
of
$\chi^2$,
$\chi_{red}^2$
and 
$Q$, where 
$k$ stands for the number of parameters.
}
\label{chi2valueflatmnras}
\begin{center}
\resizebox{12cm}{!}
{
\begin{tabular}{|c|c|c|c|c|c|c|}
\hline
Cosmology  &  SNs& $k$    &   parameters    & $\chi2$& $\chi_{red}^2$
&
$Q$      \\
\hline
\hline
flat-FLRW & 580 &  2
& $H_0=69.77\pm 0.33$; $\om= 0.295\pm0.008$
& 562.55 &  0.9732  & 0.66 \\
flat-FLRW-1 & 580 &  1
& $H_0=70$; $\om= 0.295\pm0.008$
& 563.52 &  0.9732  & 0.669 \\
$\Lambda$CDM & 580 &  3
& $H_0$ = 69.81; $\om=0.239$; $\ola=0.651$
& 562.61 &  0.975  & 0.658 \\
\hline
\end{tabular}
}
\end{center}
\end{table}

\begin{table}[ht!]
\caption
{
Numerical values 
from  the  JLA  compilation
of
$\chi^2$,
$\chi_{red}^2$
and 
$Q$, where 
$k$ stands for the number of parameters.
}
\label{chi2valueflatjla}
\begin{center}
\resizebox{12cm}{!}
{
\begin{tabular}{|c|c|c|c|c|c|c|}
\hline
Cosmology  &  SNs& $k$    &   parameters    & $\chi2$& $\chi_{red}^2$
&
$Q$      \\
\hline
flat-FLRW & 740 &  2
& $H_0=69.65\pm 0.231$; $\om= 0.3\pm0.003$
& 627.91 &   0.85  &  0.998 \\
\hline
\end{tabular}
}
\end{center}
\end{table}
As an example the luminosity distance  for the Union 2.1 
compilation 
with data as  in the first line of Table 
\ref{chi2valueflatmnras}
is 
\begin{eqnarray}
\dl(z) =
 8147.04\, \left(  1.0+z \right) \times 
\nonumber \\
   \Big  ( - 0.637664\,{\it 
F}   (  2.63214\,{\frac {\sqrt { 3.1188+
 1.33542\,z}}{ 4.64846+z}}, 0.965925 ) 
\nonumber\\
+ 1.75322
 \Big ) ~ Mpc 
\label{ldistancetutto} 
\\
when ~0 < z< 1.5
\nonumber  
\end{eqnarray}
and the distance modulus is 
\begin{eqnarray}
m-M = 
25.0 
+ 2.17147\,\ln    (  8147.04\,   (  1.0+z   )\times
\nonumber\\ 
  \Big  ( - 0.63766 {\it F}   (  2.6321 {\frac {
\sqrt { 3.1188+ 1.3354 z}}{ 4.6484+z}}, 0.96592
   ) + 1.7532   ) \Big   )
\quad.
\label{modulustutto} 
\\
when ~0 < z< 1.5
\nonumber
\end{eqnarray}

We now derive some approximate results 
without 
Legendre integral 
for the flat-FLRW case and Union 2.1  compilation
with data as in Table \ref{chi2valueflatmnras}, first line.
A Taylor  expansion of order 6 around $z$=0 of  the luminosity distance 
as  given by Eq.~(\ref{distlumflatnew}) 
for the flat-FLRW case and Union 2.1  compilation
gives 
\begin{eqnarray}
\dl(z)= 
0.000423646+ 4296.57\,z
\nonumber \\
+ 3344.13\,{z}^{2}- 1186.94\,{
z}^{3}
+ 979.403\,{z}^{4}- 42078.6\,{z}^{5} \quad Mpc
\\
when ~0 < z< 0.197
\quad.
\nonumber  
\end{eqnarray}
The upper limit in redshift, 0.197, is the value for  which 
the percentage error , see equation (\ref{error100}),
is $\delta=1.16\%$.
The asymptotic expansion of the luminosity distance 
with respect to the variable $z$  to order 5 
for the flat-FLRW case and Union 2.1  compilation
gives  
\begin{eqnarray}
\dl(z) \sim  
14283.5\,z- 15802\,{\frac {1}{\sqrt {{z}^{-1}}}}
\nonumber \\
+
 14283.5- 7901.01\,\sqrt {{z}^{-1}}+ 1975.25\, \left( {z}^
{-1} \right) ^{3/2}
~ Mpc 
\\
when ~1.27 < z< 1.5
\quad.
\nonumber  
\end{eqnarray}
At the lower limit of $z=1.27$ the percentage 
error  is $\delta=0.54\%$.  
The  two above approximations at low and  high
redshift have a limited range of existence but does 
not contain the Legendre integral 
as solutions (\ref{ldistancetutto}) and (\ref{modulustutto})
which cover the overall range $0<z<1.5$.

\begin{figure}
\includegraphics[width=10cm]{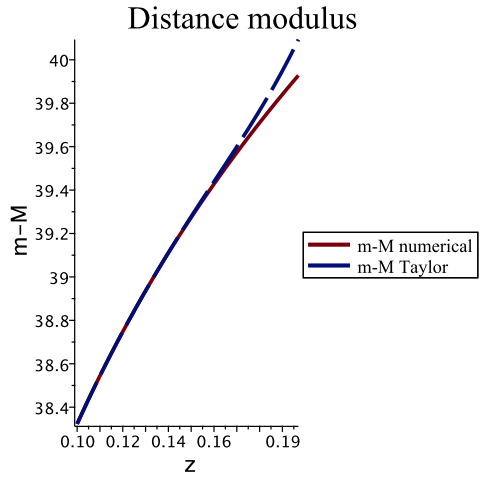}
\caption
{
Distance modulus  in flat-FLRW cosmology 
as represented by Eq.~(\ref{distmodflatnew}) with
parameters as in first line of Table 
\ref{chi2valueflatmnras} (full red line) 
and  
Taylor solution (dash-dot-dash line) (blue line).
}
\label{modulustaylor}
\end{figure}

A Taylor  expansion of order 6 
of  the distance modulus  
as  given by Eq.~(\ref{distmodflatnew})
around $z=0.1$
for the flat-FLRW case  and Union 2.1  compilation
gives 
\begin{eqnarray}
(m-M)=
36.0051+ 23.1777\,z- 109.604\, \left( z- 0.1 \right) ^{2}
+ \nonumber \\ 724.464\, \left( z- 0.1 \right) ^{3}
- 5429.06\, \left( z-
 0.1 \right) ^{4}+ 43429.8\, \left( z- 0.1 \right) ^{5}
\\
when ~0.1  < z< 0.197
\quad.
\nonumber  
\end{eqnarray}
The upper limit in redshift, 0.197, is the value at which 
the percentage error 
is $\delta=0.14\%$.
Figure \ref{modulustaylor} reports 
both the numerical   and the Taylor expansion of distance modulus 
in the above range.

The asymptotic expansion of the  distance  modulus 
with respect to the variable $z$  to order 5 
for the flat-FLRW case and Union 2.1  compilation
gives  
\begin{eqnarray}
(m-M)  \sim  
45.7741+ 2.17147\,\ln  \left( z \right) - 2.40231\,\sqrt 
{{z}^{-1}}+ 0.842625\,{z}^{-1}
\nonumber \\
+ 0.221081\, \left( {z}^{-1}
 \right) ^{3/2}- 0.570086\,{z}^{-2}- 0.150471\, \left( {z}^{-1
} \right) ^{5/2}+ 
\nonumber
\\
0.357849\,{z}^{-3}
+ 0.491842\, \left( {z}^{-
1} \right) ^{7/2}+ 0.179989\,{z}^{-4}
\\
when ~1.27 < z< 1.5
\quad.
\nonumber  
\end{eqnarray}
The lower  limit in redshift, 1.27, is the value at which 
the percentage error 
is $\delta=0.54\%$.
The ranges of existence in $z$  for the analytical approximations 
here  derived  
have the percentage error  $<~2\%$, see 
Eq.~(\ref{error100}). 

\begin{figure}
\includegraphics[width=10cm,angle=-90]{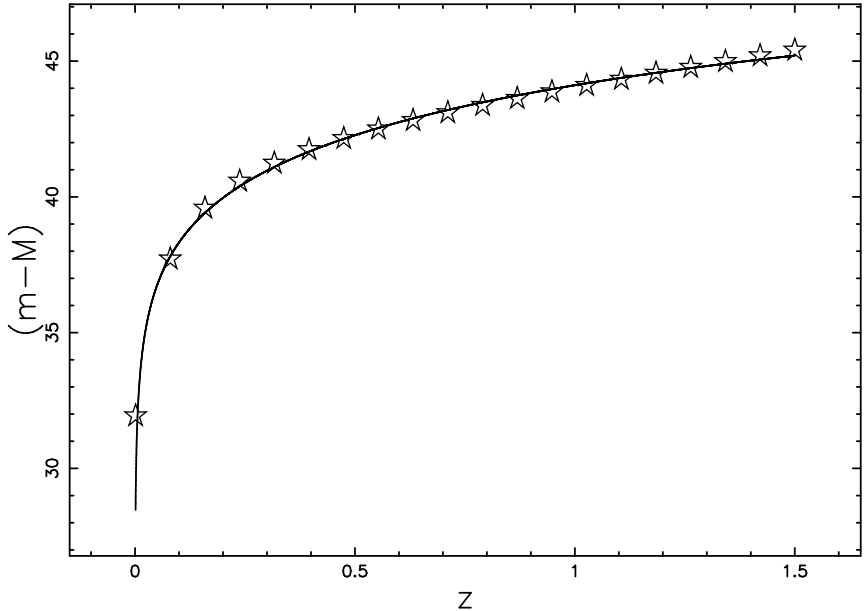}
\caption
{
Distance modulus  in flat-FLRW cosmology 
as represented by Eq.~(\ref{distmodflatnew}) with
parameters as in first line of Table 
\ref{chi2valueflatmnras} (full line) 
and  minimax rational approximation (empty stars);
Union 2.1  compilation
}
\label{distmminimax}
\end{figure}

We now introduce  the best minimax rational approximation,
see \cite{Remez1934,Remez1957,NIST2010},
of degree (2, 1) , for the distance modulus  $m_{2,1}(z) $, 
 \begin{equation}
m_{2,1}(z)  = \frac
{
a+b\,z +c\,{z}^{2} 
}
{
d+ e\,z
}
\quad .
\end{equation}
In the case in which the 
distance modulus 
is represented by Eq.~(\ref{modulustutto})
and given 
the interval $[0.001, 1.5]$,
the coefficients of the 
best minimax rational approximation
are presented in Table \ref{minimaxtable};
the maximum error for  the fit is $\approx 2.2\,10^{-5}$.
Figure \ref{distmminimax}  
displays   the data and the fit.
\begin{table}[ht!]
\caption
{
Maximum  error and 
coefficients of the distance modulus for  
the
best minimax rational approximation for the flat-FLRW case
and Union 2.1  compilation.
Interval of existence $[0.001, 1.5]$.
}
\label{minimaxtable}
\begin{center}
{
\begin{tabular}{|c|c|}
\hline
Name       &  value  \\         \hline
 maximum~error &     2.28881836\,$10^{-5}$ \\  \hline
 a         &    0.981622279  \\ \hline
 b         &    19.6473351    \\ \hline
 c         &    1.08210218    \\ \hline
 d         &     0.0309164915 \\ \hline
 e         &     0.462291896  \\ \hline
\end{tabular}
}
\end{center}
\end{table}

\section{Conclusions}

\begin{figure}
\includegraphics[width=10cm,angle=-90]{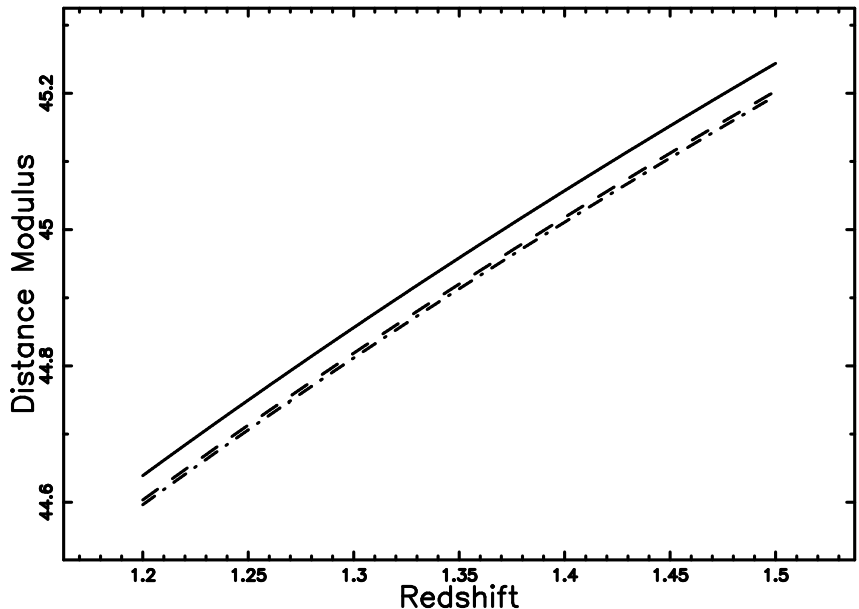}
\caption
{
Distance modulus for  
$\Lambda$CDM cosmology ( full line), 
flat-FLRW-1            ( dot-dash-dot-dash line)
and flat-FLRW cosmology( dashed line).
Parameters as in  Table 
\ref{chi2valueflatmnras}
and interval of existence $[1.2,1.5]$. 
}
\label{tresols}
\end{figure}

We have presented an analytical approximation
for the luminosity distance in terms of elliptical
integrals with real  argument.
The fit of the distance modulus of SNs of type Ia allows
finding the parameters $H_0$ and $\om$
for the two compilations in flat-FLRW cosmology
\begin{eqnarray}
H_0 =  (69.77\pm 0.33) \,  \h0units \quad \om = 0.295\pm 0.008
  \\
~~~~~~~~~ flat-FLRW-Union~2.1 
\quad ,
\nonumber \\
H_0= (69.65\pm 0.23)  \, \h0units \quad  \om= 0.3~\pm 0.003
  \\
~~~~~~~~~~~~~~ flat-FLRW-JLA 
\quad .
\nonumber
\end{eqnarray}
A first  comparison with \cite{Benitez-Herrera2013}
in the case of the Union~2.1 compilation 
gives a percentage error $p=0.93\%$ for the 
derivation of $H_0$ and 
$p=0.67\%$ for the 
derivation of $\om$.
A second  comparison can be done with  equation (13) in 
\cite{Planck2018}
\begin{eqnarray}
H_0 = (67.27 \pm 0.60) \,  \h0units \quad \om = 0.3166 \pm
0.0084
 \\
~~~~~~~~~~~~~~~Planck~2018\quad .
\nonumber 
\end{eqnarray}
In the case of the Union~2.1 compilation, 
the  percentage error $p=3.71\%$ for the 
derivation of $H_0$ and 
$p=6.82\%$ for $\om$. 
A Taylor expansion at low redshift and 
an asymptotic expansion are presented both
for the luminosity distance and the distance modulus.
A simple version of the distance 
modulus is determined 
through the 
best minimax rational approximation.
Adopting the cosmological 
parameters  found here, the cosmological constant $\Lambda$ turns out to
be, for the Union 2.1 compilation,
\begin{eqnarray}
\Lambda=
(1.19457 \pm 0.017)10^{-52} \frac{1}{m^2}
  \\
~~~~~~~~~~~~~~~~~~~ flat-FLRW~ Union~2.1 \quad ,
\nonumber 
\end{eqnarray} 
or introducing $c=1$ 
and the  Planck time, $t_p$,
\begin{eqnarray}
\Lambda=
(3.12046 \pm 0.0462942)10^{-122} \frac{1}{t_p^2}
 \\
~~~~~~~~~~~~~~~~~flat-FLRW~ Union~2.1 \quad .
\nonumber
\end{eqnarray} 
The statistical parameters of the fits are given 
in Tables \ref {chi2valueflatmnras} and 
\ref{chi2valueflatjla} where the other two models are 
presented.
The values  of the 
$\chi^2$ in the above table  say  that for  the Union 2.1 compilation
the flat cosmology  produces a better fit 
than  the $\Lambda$CDM does, but 
the situation is the reverse for the JLA compilation.
As a concluding remark we  point out that,
thanks to the  
calibration on the distance modulus
of SNs,  the differences between the
solutions here analyzed are minimum.
Therefore   a restricted range in redshift should be adopted 
in order to visualize the diverseness,
see Figure \ref{tresols}.

\providecommand{\newblock}{}

\providecommand{\newblock}{}
\end{document}